\documentclass[conference]{IEEEtran}
\IEEEoverridecommandlockouts
\usepackage{cite}
\usepackage{amsmath,amssymb,amsfonts,bm}
\usepackage{algorithmic}
\usepackage{graphicx}
\usepackage{textcomp}
\usepackage{xcolor}

\usepackage{algorithm}
\usepackage{subcaption}
\usepackage{hyphenat}
\usepackage{booktabs}
\usepackage{siunitx}
\usepackage{url}
\usepackage{hyperref}

\usepackage[top=0.75in, left=0.63in, right=0.63in, bottom=1.04in]{geometry}
\newcommand\rurl[1]{%
  \href{https://#1}{\nolinkurl{#1}}%
}

\def\BibTeX{{\rm B\kern-.05em{\sc i\kern-.025em b}\kern-.08em
    T\kern-.1667em\lower.7ex\hbox{E}\kern-.125emX}}
\begin{document}

\title{Accent Conversion in Text-To-Speech Using Multi-Level VAE and Adversarial Training
\thanks{This project has received funding from SUTD Kickstarter Initiative no. SKI 2021\_04\_06.}
}

\author{
    \IEEEauthorblockN{Jan Melechovsky\IEEEauthorrefmark{1}, Ambuj Mehrish\IEEEauthorrefmark{1},
    Berrak Sisman\IEEEauthorrefmark{2}, and Dorien Herremans\IEEEauthorrefmark{1}
    }
    \IEEEauthorblockA{
        \IEEEauthorrefmark{1}Audio, Music, and AI Lab, Singapore University of Technology and Design, Singapore\\
        jan\_melechovsky@mymail.sutd.edu.sg, ambuj\_mehrish@sutd.edu.sg, dorien\_herremans@sutd.edu.sg
    }
    \IEEEauthorblockA{
        \IEEEauthorrefmark{2}Speech \& Machine Learning Lab, The University of Texas at Dallas, USA\\
        Berrak.Sisman@UTDallas.edu
    }
}


\maketitle

\begin{abstract}
With rapid globalization, the need to build inclusive and representative speech technology cannot be overstated. Accent is an important aspect of speech that needs to be taken into consideration while building inclusive speech synthesizers.
Inclusive speech technology aims to erase any biases towards specific groups, such as people of certain accent. We note that state-of-the-art Text-to-Speech (TTS) systems may currently not be suitable for all people, regardless of their background, as they are designed to generate high-quality voices without focusing on accent.
In this paper, we propose a TTS model that utilizes a Multi-Level Variational Autoencoder with adversarial learning to address accented speech synthesis and conversion in TTS, with a vision for more inclusive systems in the future. We evaluate the performance through both objective metrics and subjective listening tests. The results show an improvement in accent conversion ability compared to the baseline.
\end{abstract}


\section{Introduction}
\label{sec:intro}




In a globalized world, we encounter speakers of various accents on a daily basis, yet there is still a lack of representation of all these accents in speech technology. Accent is an important attribute that influences mutual intelligibility of a spoken dialogue. It refers to a peculiar way of speaking a language that can be described on phonemic, phonetic, rhythmical, and structural levels~\cite{wells1982accents}. As part of one's idiolect, accent forms a part of a person's identity. 
Given the importance of accent in speech synthesis, paradoxically, there has not been much research done in this field when it comes to TTS.

\begin{figure*}[h]
    \centering
    \includegraphics[width=2\columnwidth]{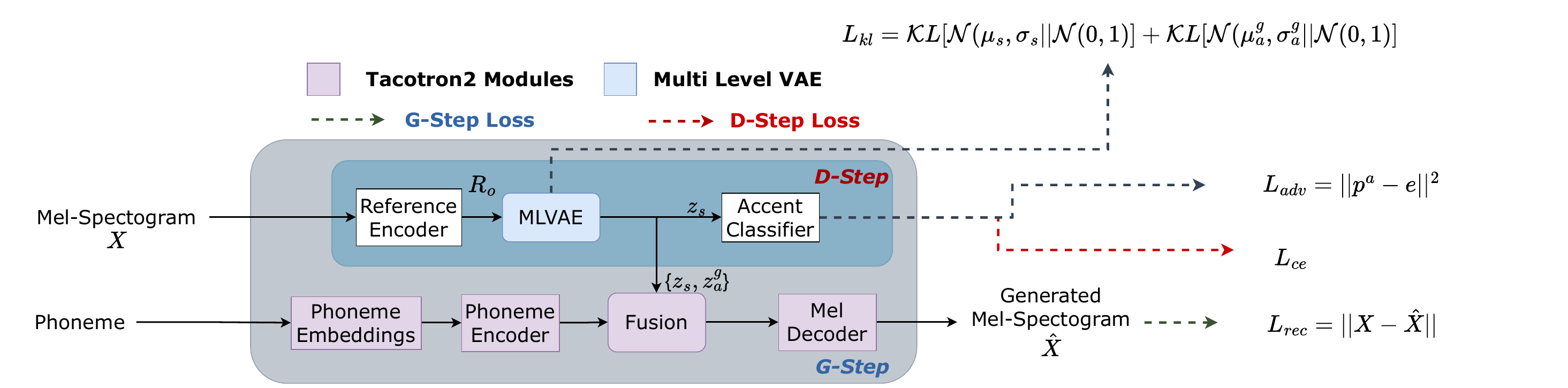}
    \caption{The proposed model architecture with D-Step (discriminator) and G-Step (generator) illustrations.}
    \label{fig:tts}
    \vspace{-5mm}
\end{figure*}

The task of converting one's accent into another is a domain of Accent Conversion (AC), or its more specific variant -- Foreign Accent Conversion (FAC) \cite{wang2021accent,DING2022101302}. In FAC, the input audio of an L2 (second language) speaker is converted to an output audio of the same speaker with an L1 (native) accent. One of the main challenges in FAC is the disentanglement of speaker identity and accent, which is crucial to achieving good conversion, but challenging since the two overlap substantially. The main application area of FAC is Computer-Assisted Pronunciation Training \cite{rogerson2021computer}, which aims to help L2 speakers' pronunciation to be closer to L1.
Our work, however, puts emphasis on foreign accents, as they are scarcely present in current speech synthesis systems. As in FAC, we aim to keep the original speaker's identity, while changing the speaker's accent into one of many other accents, which might in fact change the person's perceived identity accent-wise. 



In controllable TTS, studies have focused on modelling speaker attributes such as pitch, pace, and intonation, e.g. in \cite{wang2018style} or \cite{hsu2018hierarchical}, where Tacotron models are enhanced with modules of Global Style Tokens (GST), and Gaussian Mixture Variational Autoencoder, respectively. When it comes to modelling accent in TTS specifically, one of the first papers is by Melechovsky et al. \cite{melechovsky2023learning}.
They proposed a Tacotron2 model enhanced with a Multi-Level Variational Autoencoder (MLVAE) to capture speaker characteristics and accent. This framework showed promising results, but seemed to lack a bit in its accent conversion strength. Since the change of accent is entirely dependent on the accent embeddings, we hypothesize that putting more weight on these accent embeddings might strengthen the accent conversion ability. In an FAC model by Wang et al. \cite{wang2021accent}, adversarial learning is adopted to wipe out speaker-related information from the extracted speech recognition bottleneck features before passing them on to the conversion model. Inspired by this, we believe adversarial loss can be used to enhance the accent conversion ability of the MLVAE model.

In this paper, we propose an MLVAE-based accented TTS system with a two-step training procedure based on adversarial loss. The purpose of adversarial learning is to wipe out accent information from the speaker embeddings so as to lessen the dominance of speaker identity during inference. This better disentanglement of the speaker from the accent representation may result in a potentially stronger accent conversion. The contributions of this paper are as follows: 1)~we propose a novel MLVAE-based Tacotron2 system for accented TTS that utilizes adversarial learning; 
2)~the proposed method shows stronger accent conversion ability than its predecessor; and 
3)~we discuss the shortcomings of current accented TTS research and propose future directions.

The rest of the paper is organized as follows: In Section~\ref{sec:related}, related work is presented. Section \ref{sec:proposed} dives into our proposed method, and in Sections~\ref{sec:results} and~\ref{sec:conc} we present the results, discuss their meaning, and touch on the limitations of this work and research field.

\section{Related Work}
\label{sec:related}
Few studies have focused on capturing accents in TTS. In \cite{hsu2018hierarchical}, a Gaussian Mixture VAE was used to control speech attributes, including accent, though this was not the primary focus of the model. Liu et al. \cite{liu2024controllable} tackled accent modification by adjusting phoneme energy, pitch, and duration through an accent variance adaptor. Melechovsky et al. \cite{melechovsky2023learning} specifically addressed accented TTS using an MLVAE module to disentangle speaker and accent information, producing robust, speaker-independent accent embeddings. The model included an accent classifier to cluster these embeddings further, but its impact was not demonstrated.
However, these accent embeddings are not forced to be utilized by the framework to reconstruct the audio, as sole speaker embeddings are sufficient as they can still contain original accent information, which results in a poor accent conversion performance, hence we aim to improve the accent conversion ability through adversarial learning, as proposed in the next section.

\section{Proposed Method}
\label{sec:proposed}

In this section, we formulate the accent conversion problem for TTS, with a two-step training procedure based on adversarial loss as shown in Fig. \ref{fig:tts}. The proposed approach is motivated by the fact that accent and speaker identity are highly correlated attributes of speech, and the MLVAE model can, to a large extent, learn both the attributes in one of the two designated embeddings only. Therefore, we propose to utilize an accent classifier with adversarial loss to increase disentanglement and minimize the accent information present in the speaker representation to strengthen the model's accent conversion ability. We refer to our model as MLVAE-ADV. 


 \begin{figure}[h]
    \centering
    \includegraphics[width=1.1\columnwidth]{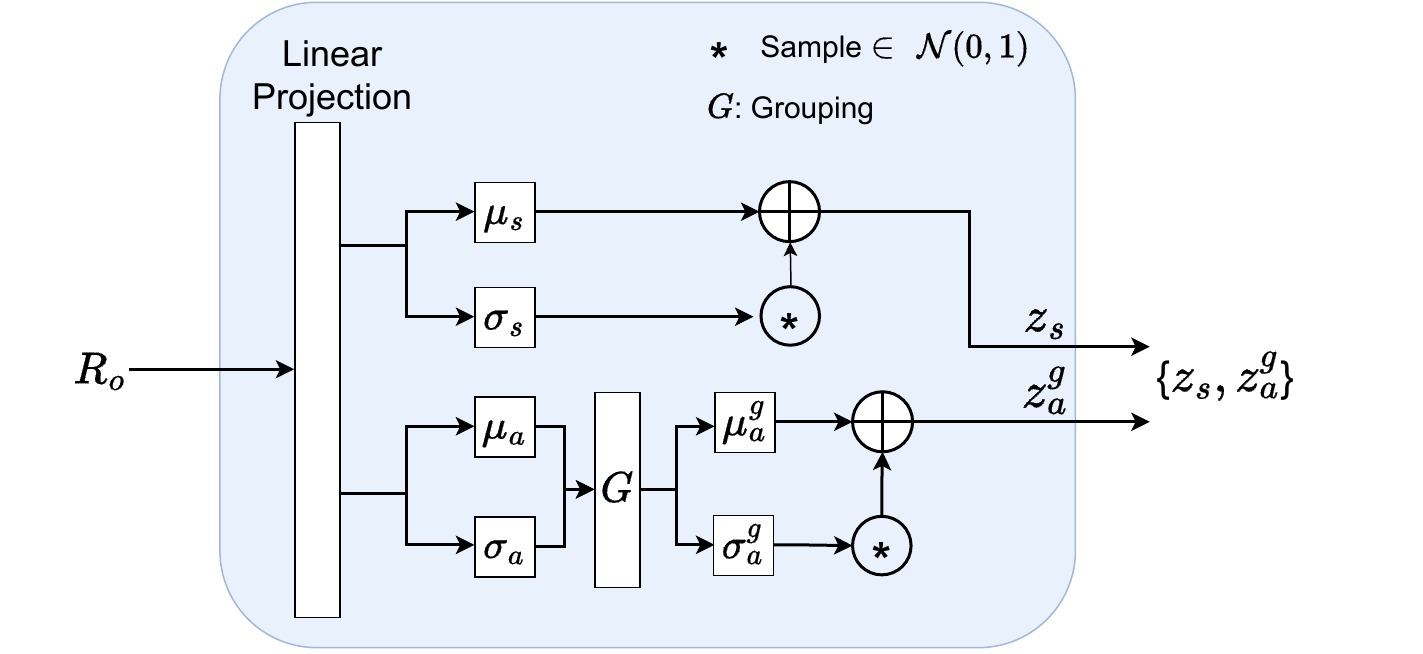}
    \caption{Detailed view of the MLVAE encoder, $R_o$ is the output of the Reference Encoder.}
    \label{fig:mlvae}
\end{figure}
\subsection{MLVAE Encoder}
The Multi-Level Variational Autoencoder (ML-VAE) architecture \cite{bouchacourt2018multi} is used for encoding disentangled representations of a set of grouped observations based on different accents, as illustrated in Fig.~\ref{fig:mlvae} with Tacotron2 \cite{shen2018natural} as the base model. The accents are considered as a variation factor between different speakers. To disentangle accent and speaker attributes in the latent representation, two latent variables are used: $z_s$ and $z_a^g$, where $z_a^g$ represents the factor of variation among the accents $a$, superscript $g$ represents the grouping, and  $s$ represents speaker. The observations are assumed to be independent and identically distributed (i.i.d) at the group level, and the average marginal log-likelihood over different accents is defined as $\mathcal{A}$ as $\frac{1}{|\mathcal{A}|} \sum_{a \in \mathcal{A}} \log p(X_{a}|\theta)$, where $\theta$ are the network parameters.

The marginal log-likelihood $\log p(X_{a}|\theta)$ can be expressed in terms of Kullback-Leibler (KL) divergence between the true posterior $p(z_{a}^{g},z_{s}|X_{a};\theta)$ and the variational approximation $q(z_{a}^{g},z_{s}|X_{a};\phi_{a})$. The group Evidence Lower Bound $ELBO(a;\theta,\phi_{s},\phi_{a})$ is also defined.
 \begin{equation}
\label{elbo}
\begin{split}
        \log p(X_{a}|\theta) = & ELBO(a;\theta,\phi_{s},\phi_{a}) +  \\ & KL(q(z_{a}^{g},z_{s}|X_{a};\phi_{a})||p(z_{a}^{g},z_{s}|X_{a};\theta)) \\
        \geq & ELBO(a;\theta,\phi_{s},\phi_{a})
\end{split}
\end{equation}

\begin{figure*}
    \centering
    \begin{subfigure}[t]{0.3\textwidth}
        \includegraphics[width=\columnwidth,height=0.5\columnwidth]{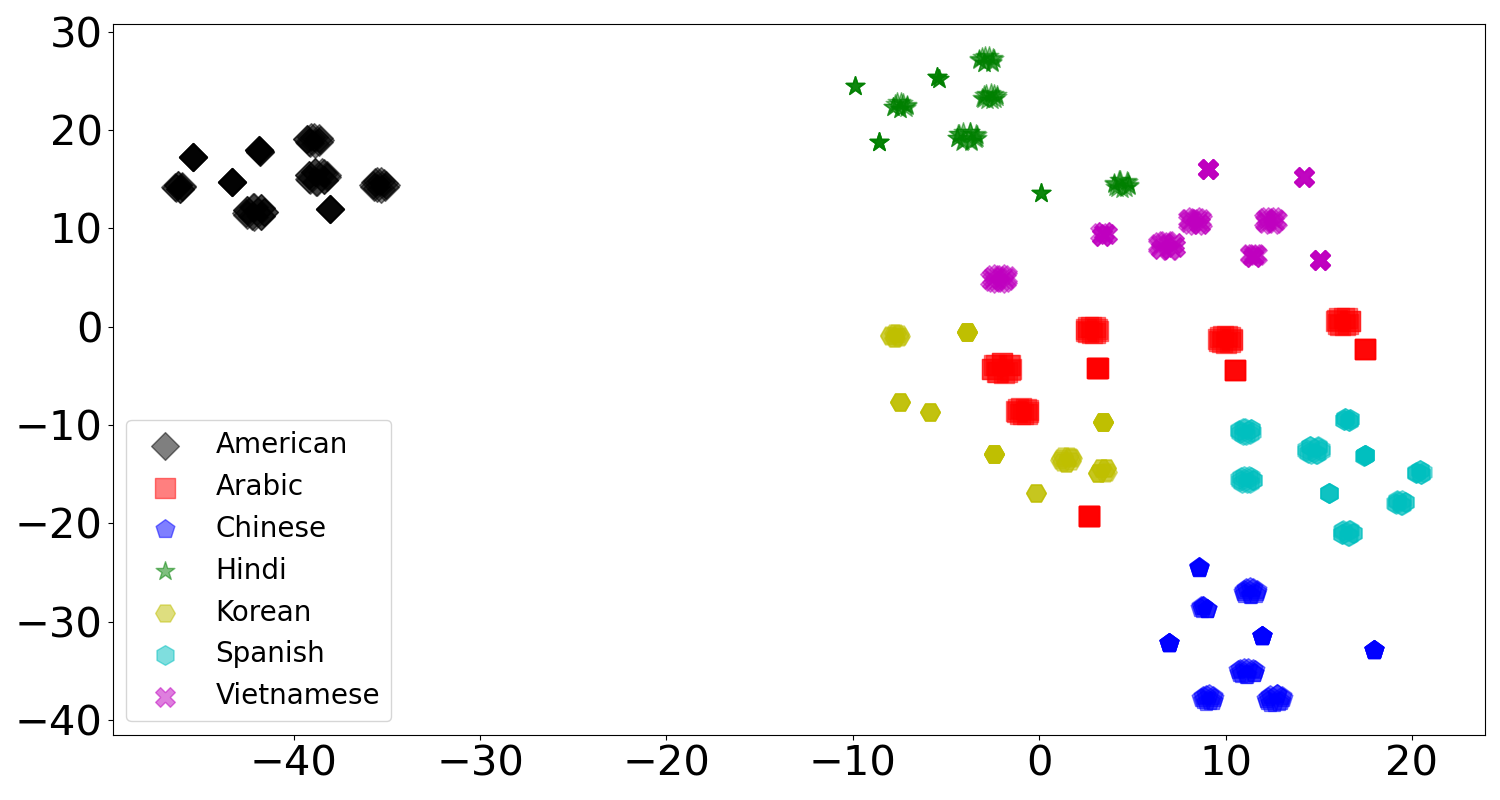} 
        \caption{Grouped accent embeddings $z_a$.}
        \label{fig:acc_embs} 
    \end{subfigure}
    ~
    \begin{subfigure}[t]{0.3\textwidth}
        \includegraphics[width=\columnwidth,height=0.5\columnwidth]{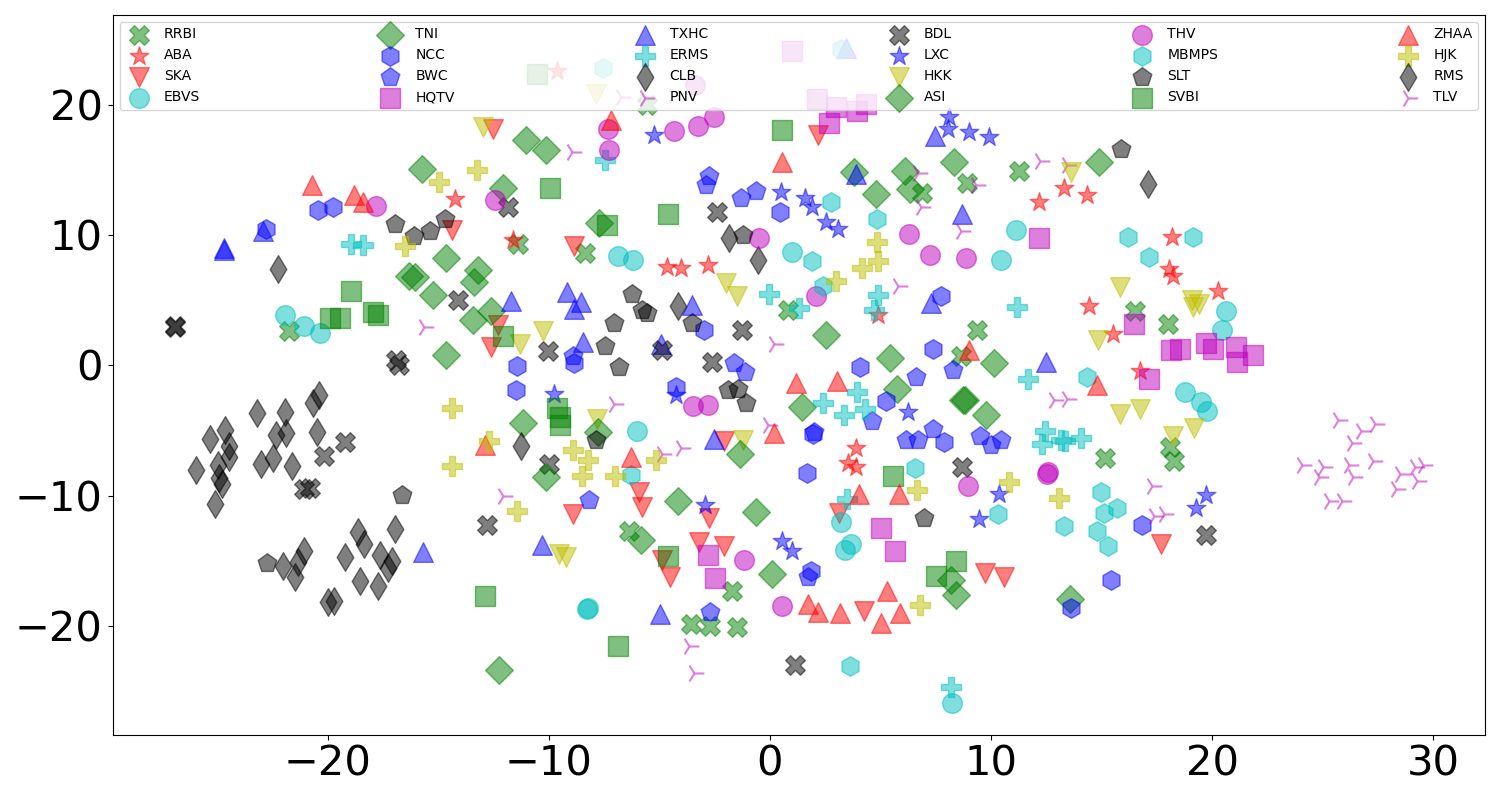}
        \caption{Speaker embeddings with $\gamma$ too high.}
        \label{fig:spk_embs_bad}
    \end{subfigure}
    ~
    \begin{subfigure}[t]{0.3\textwidth}
        \includegraphics[width=\columnwidth,height=0.5\columnwidth]{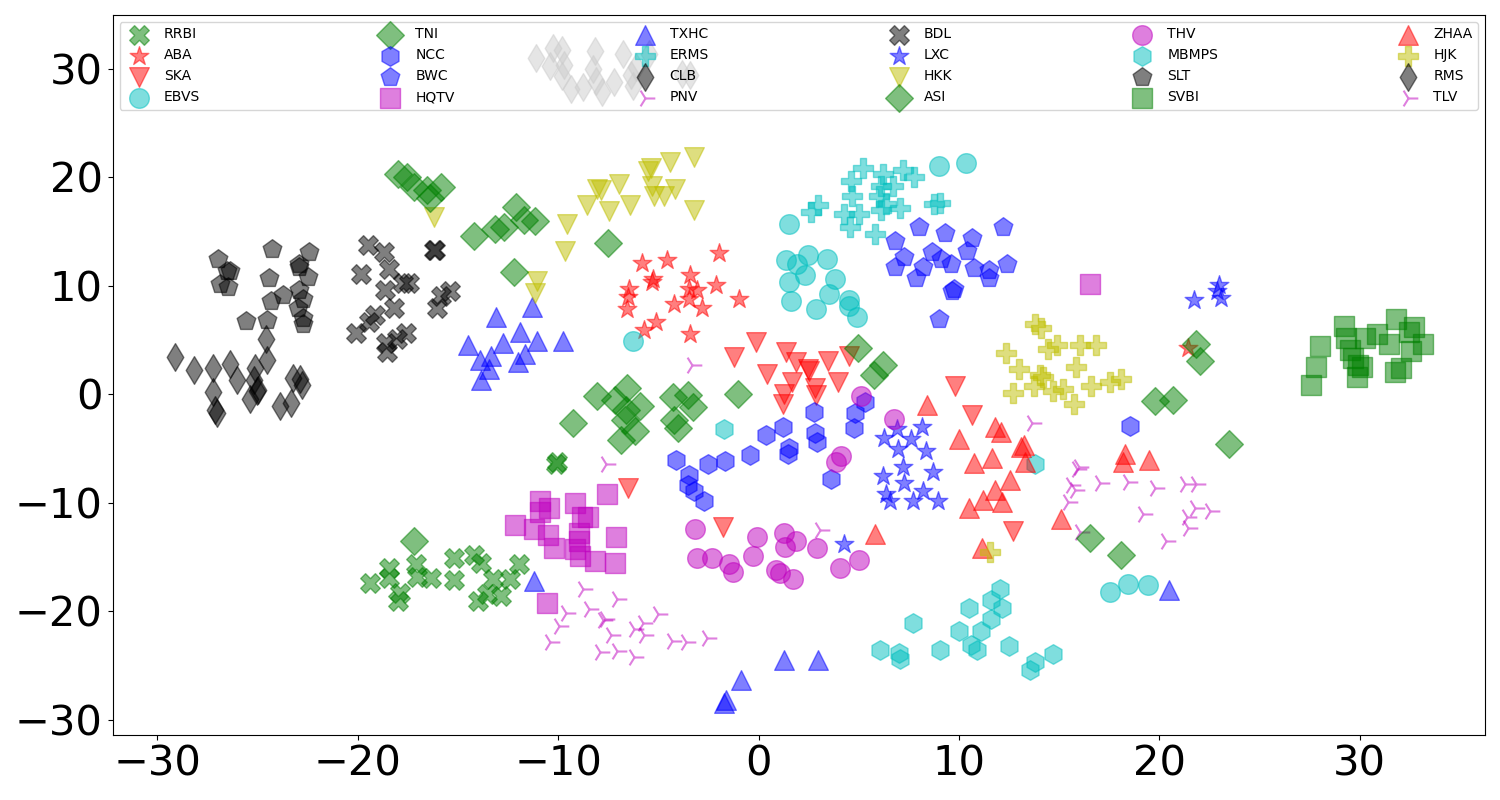}
        \caption{Speaker embeddings with optimal $\gamma$.}
        \label{fig:spk_embs_optimal}
    \end{subfigure}
    \caption{A t-SNE projection of speaker and accent embeddings from the MLVAE-ADV model.
    }
     \vspace{-3mm}
\end{figure*}

$ELBO$ is a lower bound on the marginal log-likelihood, and the KL divergence is always positive. The variational parameters $\phi_{s}$ and $\phi_{a}$ are utilized in this context. The traditional VAE learns the model parameters by maximizing the average $ELBO$ over individual samples, while the ML-VAE maximizes the group $ELBO$ using mini-batches of the group. We can observe the effect of grouping on accent embeddings in Fig. \ref{fig:acc_embs}. In this work, we maximize the average group $ELBO$ to train the model parameters, as discussed in \cite{bouchacourt2018multi}. The KL loss $\mathcal{L}_{kl}$ is computed as illustrated in Fig. \ref{fig:tts}, and the reconstruction loss between the predicted mel spectrogram $\hat{X}$ and the ground truth mel spectrogram $X$ is computed as follows:
\begin{equation}
    L_{recon} = ||\hat{X}-X||_{2}
\end{equation}
where $||.||_{2}$  denotes $L_{2}$ norm. During the $G$-Step (generator), the model is trained to generate mel spectograms ($\hat{X}$) from phoneme sequences~($P$) and the reference spectrograms ($X$), and all the model parameters are updated. In the $D$-Step (discriminator), we update only the accent classifier. The detailed training procedure is outlined in Algorithm \ref{trainalgo}.

\begin{algorithm}
    \caption{Adversarial Training}\label{trainalgo}
    \begin{algorithmic}[1]
        \STATE \%\% $P$: Phoneme Embeddings \%\%
        \STATE \%\% $\hat{y}$: accent labels \%\%
        \STATE $\mathcal{G} :=$  \{phoneme\_encoder (pe), mel\_decoder (md), reference\_encoder (re), MLVAE, accent\_classifier (ac)\}
        \STATE $\mathcal{D}:=$ \{reference\_encoder (re), MLVAE, accent\_classifier (ac)\}
        \FOR{epochs = $1$ to $N$}
            \STATE Sample {$\{X^{(i)}\}^{m}_{i=1} \sim \mathcal{P}_{r}$} a batch from the real data
            \STATE require\_grad: $\mathcal{G} \backslash \{ac\}$ $\rightarrow$ True
              \STATE $G$-Step:

                    \STATE \hskip1.5em Compute the output $\hat{X^{i}} = \mathcal{G}(X^{i},P)$
                    \STATE \hskip1.5em Compute $\mathcal{L}_{kl}$, $\mathcal{L}_{adv}$, $\mathcal{L}_{rec}$
                    \STATE \hskip1.5em Update $\theta^{(\mathcal{G})} \leftarrow \theta^{(\mathcal{G})} + \eta \frac{d\mathcal{L} \theta^{\mathcal{G}}}{d\theta^{\mathcal{G}}}$ 

             \STATE require\_grad: $\mathcal{G}$ $\rightarrow$ False, $\mathcal{D} \backslash \{ac\}$ $\rightarrow$ False
            \STATE $D$-Step:
                   \STATE \hskip1.5em Compute the output $\hat{y} = \mathcal{D}(X^{i})$
                   \STATE \hskip1.5em Compute  $\mathcal{L}_{ce}$
                   \STATE \hskip1.5em Update $\theta^{(\mathcal{D})}_{ac} \leftarrow \theta^{(\mathcal{D})}_{ac} + \eta \frac{d\mathcal{L} \theta^{\mathcal{D}}_{ac}}{d\theta^{\mathcal{D}}_{ac}}$ 
        \ENDFOR
    \end{algorithmic}
\end{algorithm}

To identify and disentangle accent-related information from $z_s$, we utilize the adversarial accent classifier, as we incorporate an adversarial loss function, defined as follows:

\begin{equation}
    L_{adv} = ||e-\hat{p}^{a}||^{2}_{2}
\end{equation}

We define $e$ as a uniform distribution vector, where each element is equal to $1/\lvert A \rvert$, $\lvert A \rvert$ being the total number of accents in the set $\mathcal{A}$; and $\hat{p}^{a}$ represents the output probability of the adversarial accent classifier.
The total loss for training the model during the $G$-Step can be computed as follows:

\begin{equation}
    \mathcal{L}_{G} = \mathcal{L}_{recon} + \beta \mathcal{L}_{kl} + \gamma \mathcal{L}_{adv}
\end{equation}

where $\beta$ and $\gamma$ are coefficients associated with the KL and adversarial loss. Finally, $\mathcal{L}_{D} = \alpha \mathcal{L}_{ce}$ is used during the $D$-Step, where $\alpha$ is the coefficient associated with cross entropy loss. The tuning of hyperparameters $\alpha$, $\beta$ and, $\gamma$ is further explained in Section \ref{sec:trainparam}.

\section{Experiments and results}
\label{sec:results}
\subsection{Dataset and Baselines}

In our experiments, we used two datasets: L2Arctic and CMUArctic \cite{zhao2018l2,kominek2004cmu}. L2Arctic includes 24 speakers (balanced by sex) with 6 accents (Arabic, Chinese, Hindi, Korean, Spanish, and Vietnamese), each represented by 4 speakers. We added 4 American-accented speakers (BDL, CLB, RMS, and SLT) from CMUArctic. Our dataset comprised 28 speakers, 7 accents, and approximately 31 hours of data. We used 10 unseen utterances per speaker for testing, 20 for validation, and the rest for training.

For our first baseline, we chose GST \cite{wang2018style} with an embedding size of $256$, $10$ tokens, and $8$ attention heads; combined with the same Tacotron2 model we use in our proposed model. We note that GST is unsupervised and has not been modified for accent conversion, thus we use it without converting accent for non-conversion-related comparison. While it is possible to naively interpolate between a source speaker and an average of target accent speakers in their embedding space, it would produce speech with a new speaker identity. Our second baseline is the MLVAE model from \cite{melechovsky2023learning}, without their originally proposed accent classifier.

\subsection{Experimental Setup and Inference}
\label{sec:trainparam}
All models were trained with a batch size of 64 for 200k steps and ADAM optimizer.
Both MLVAE models had accent and speaker embedding sizes of 128. The $\beta$ coefficient started at $10^{-6}$ and increased to $10^{-4}$ between 5k and 15k steps to aid training. The MLVAE-ADV model used $\alpha$ of $10^{-1}$ for the accent classifier in D-step and $\gamma$ of $10^{-2}$ throughout training. We experimented with $\gamma$ coefficients from $10^{-4}$ to 0.5. Low $\gamma$ resulted in no accent conversion, while high $\gamma$ (e.g., $10^{-1}$) caused chaotic speaker embeddings (Fig. \ref{fig:spk_embs_bad}) and unintelligible output speech. Optimal speaker embeddings showed visible but not isolated speaker clusters, reflecting partial identity subtraction (accent) (Fig. \ref{fig:spk_embs_optimal}).
\begin{figure}[h]
    \centering
    \includegraphics[width=1\columnwidth]{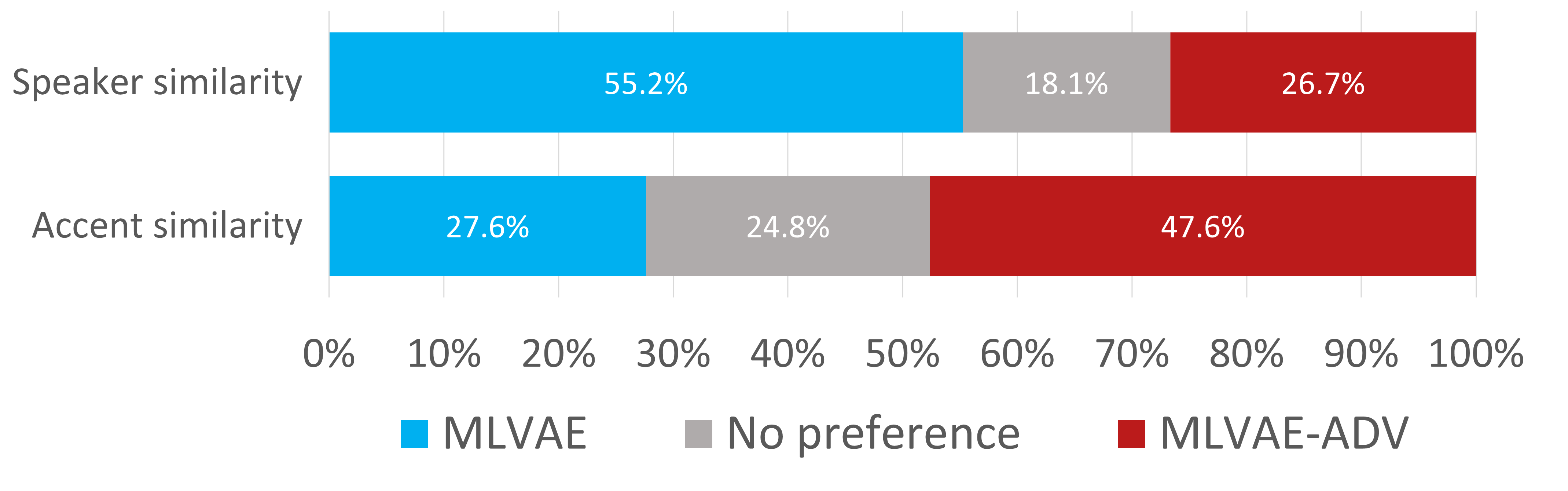}
    \caption{XAB accent and speaker similarity test results.}
    \label{fig:xab}
\end{figure}

During inference, speaker and accent embeddings can be extracted from either single or multiple reference audio files. In our study, we passed the complete validation set through the model to extract and preserve the embeddings. Subsequently, during inference, we load these embeddings and compute the average representation of the selected speaker and the target accent (reproducing the grouping operation).
A pre-trained HiFiGAN vocoder is used to process outputs into speech.

\subsection{Objective Evaluation}
\label{sec:obj_eval}
To assess the effectiveness of reconstructing the mel spectrogram and the intelligibility of the synthesized speech, we utilize mel-cepstral distortion (MCD) \cite{toda2007voice} and word error rate (WER), respectively. The results are presented in Table \ref{tab:obj_eval}. Our analysis indicates that the proposed MLVAE-ADV outperforms MLVAE and GST in mel spectrogram reconstruction ability. However, with a WER of $0.2124$, the performance of MLVAE-ADV falls short compared to $0.1668$ and $0.1647$ for GST and MLVAE, respectively. We hypothesize that the superior accent conversion capability of MLVAE-ADV could impact its lower WER compared to other baselines.

\subsection{Subjective Evaluation}
\label{sec:subj_eval}
We conducted listening experiments to further assess the performance of MLVAE-ADV. A total of 15 participants joined the study and each listened to a total of 77 samples. 
\footnote{Audio samples and source code are available at \rurl{amaai-lab.github.io/Accented-TTS-MLVAE-ADV/}}.



In the first part of the test, listeners rated the voice quality of samples from two baseline models, the proposed method, and the ground truth on a five-point Likert Scale (1: bad, 2: poor, 3: fair, 4: good, 5: excellent). The resulting mean Opinion Score (MOS) is shown in Table~\ref{tab:obj_eval}. A paired t-test shows a statistically significant difference in voice quality when comparing ground truth to all models ($p<0.001$); MLVAE-ADV to GST ($p<0.01$); and MLVAE-ADV to MLVAE ($p<0.001$).

In the second part, listeners rated accent-converted samples for speaker similarity in an XAB test. They were presented with an original speaker sample X, and samples from the MLVAE (A) and MLVAE-ADV (B) models. Listeners picked which sample (A or B) retains more of the original speaker's (X) identity. The results in Fig.~\ref{fig:xab} show that the MLVAE model is preferred over MLVAE-ADV in terms of speaker similarity, suggesting that the accent conversion in MLVAE is weaker, thus preserving the original identity better.




Finally, we test for accent similarity in an XAB test. The listeners were presented with the same set of accent-converted samples (A, B) as in the previous test, but this time, we also presented them with a target accent reference sample Y from one of the speakers of that accent. Listeners were to determine whether A or B sounds more similar in terms of accent similarity to the reference accent Y.
In this test (Fig.~\ref{fig:xab}), MLVAE-ADV outperforms the baseline of MLVAE. 



\begin{table}[]
    \centering
    \caption{Objective Evaluation results (top part; the lower, the better) and Subjective evaluation results (bottom part) for voice quality, Mean Opinion Score with 95\% Confidence Interval.}
    \begin{tabular}{@{}lcccc@{}}
    \toprule
    Metric & GT & GST & MLVAE & MLVAE-ADV\\
    \midrule
    MCD $\downarrow$ & -  &7.0576 &7.0160 &6.9422\\
    WER $\downarrow$ &0.1370 &0.1647 &0.1668 &0.2124\\
    \midrule
    MOS &4.493	&3.236	&3.273	&2.941\\
    95\% CI &0.419	&0.411	&0.323	&0.283\\ \bottomrule
    \end{tabular}
    \label{tab:obj_eval}
\end{table}
\section{Conclusion}
\label{sec:conc}

We proposed the MLVAE-ADV framework, a Tacotron2-based MLVAE model with adversarial training to improve accent conversion in accented TTS. Evaluations show increased accent conversion but some loss in speaker identity and voice quality. Fine-tuning loss coefficients may mitigate this quality decrease. Since accent is a key part of an individual's idiolect, changing it may alter perceived speaker identity. This trade-off could be due to the limited dataset, which includes only 4 speakers per accent. This work offers new directions for future research in accented TTS. Future studies should use larger datasets and better balance accent conversion with speaker identity preservation to advance inclusive speech technology.

\bibliographystyle{IEEEtran}
\bibliography{biblio}

\end{document}